\documentclass[aps,prd,superscriptaddress,showpacs,preprint,tightenlines,byrevtex]{revtex4-1}

\usepackage{graphicx}
\usepackage{amssymb,amsmath}
\usepackage{multirow}

\graphicspath{{fig/paper/}}

\newcommand{\mev}{\mbox{\ensuremath{\text{MeV}}}}

\newcommand{\gevcc}{\mbox{\ensuremath{\text{GeV}/c^2}}}

\begin{document}


\preprint{\vbox{ \hbox{ }
                        \hbox{Belle Preprint 2012-28}
                        \hbox{KEK Preprint 2012-32}
                        \hbox{arXiv: 1301.1105}
                        \hbox{Phys. Rev. D. \textbf{87}, 071102 (2013)}
}}

\title{\quad\\[0.5cm]Search for heavy neutrinos at Belle}

\affiliation{University of the Basque Country UPV/EHU, 48080 Bilbao}
\affiliation{University of Bonn, 53115 Bonn}
\affiliation{Budker Institute of Nuclear Physics SB RAS and Novosibirsk State University, Novosibirsk 630090}
\affiliation{Faculty of Mathematics and Physics, Charles University, 121 16 Prague}
\affiliation{University of Cincinnati, Cincinnati, Ohio 45221}
\affiliation{Justus-Liebig-Universit\"at Gie\ss{}en, 35392 Gie\ss{}en}
\affiliation{Gifu University, Gifu 501-1193}
\affiliation{Gyeongsang National University, Chinju 660-701}
\affiliation{Hanyang University, Seoul 133-791}
\affiliation{University of Hawaii, Honolulu, Hawaii 96822}
\affiliation{High Energy Accelerator Research Organization (KEK), Tsukuba 305-0801}
\affiliation{Hiroshima Institute of Technology, Hiroshima 731-5193}
\affiliation{Ikerbasque, 48011 Bilbao}
\affiliation{Indian Institute of Technology Guwahati, Assam 781039}
\affiliation{Indian Institute of Technology Madras, Chennai 600036}
\affiliation{Institute of High Energy Physics, Chinese Academy of Sciences, Beijing 100049}
\affiliation{Institute of High Energy Physics, Vienna 1050}
\affiliation{Institute for High Energy Physics, Protvino 142281}
\affiliation{Institute for Theoretical and Experimental Physics, Moscow 117218}
\affiliation{J. Stefan Institute, 1000 Ljubljana}
\affiliation{Kanagawa University, Yokohama 221-8686}
\affiliation{Institut f\"ur Experimentelle Kernphysik, Karlsruher Institut f\"ur Technologie, 76131 Karlsruhe}
\affiliation{Korea Institute of Science and Technology Information, Daejeon 305-806}
\affiliation{Korea University, Seoul 136-713}
\affiliation{Kyungpook National University, Daegu 702-701}
\affiliation{\'Ecole Polytechnique F\'ed\'erale de Lausanne (EPFL), Lausanne 1015}
\affiliation{Faculty of Mathematics and Physics, University of Ljubljana, 1000 Ljubljana}
\affiliation{Luther College, Decorah, Iowa 52101}
\affiliation{University of Maribor, 2000 Maribor}
\affiliation{Max-Planck-Institut f\"ur Physik, 80805 M\"unchen}
\affiliation{School of Physics, University of Melbourne, Victoria 3010}
\affiliation{Moscow Physical Engineering Institute, Moscow 115409}
\affiliation{Moscow Institute of Physics and Technology, Moscow Region 141700}
\affiliation{Graduate School of Science, Nagoya University, Nagoya 464-8602}
\affiliation{Kobayashi-Maskawa Institute, Nagoya University, Nagoya 464-8602}
\affiliation{Nara Women's University, Nara 630-8506}
\affiliation{National Central University, Chung-li 32054}
\affiliation{National United University, Miao Li 36003}
\affiliation{Department of Physics, National Taiwan University, Taipei 10617}
\affiliation{H. Niewodniczanski Institute of Nuclear Physics, Krakow 31-342}
\affiliation{Nippon Dental University, Niigata 951-8580}
\affiliation{Niigata University, Niigata 950-2181}
\affiliation{Osaka City University, Osaka 558-8585}
\affiliation{Pacific Northwest National Laboratory, Richland, Washington 99352}
\affiliation{Panjab University, Chandigarh 160014}
\affiliation{Research Center for Electron Photon Science, Tohoku University, Sendai 980-8578}
\affiliation{University of Science and Technology of China, Hefei 230026}
\affiliation{Seoul National University, Seoul 151-742}
\affiliation{Sungkyunkwan University, Suwon 440-746}
\affiliation{School of Physics, University of Sydney, NSW 2006}
\affiliation{Tata Institute of Fundamental Research, Mumbai 400005}
\affiliation{Excellence Cluster Universe, Technische Universit\"at M\"unchen, 85748 Garching}
\affiliation{Toho University, Funabashi 274-8510}
\affiliation{Tohoku Gakuin University, Tagajo 985-8537}
\affiliation{Tohoku University, Sendai 980-8578}
\affiliation{Department of Physics, University of Tokyo, Tokyo 113-0033}
\affiliation{Tokyo Institute of Technology, Tokyo 152-8550}
\affiliation{Tokyo Metropolitan University, Tokyo 192-0397}
\affiliation{Tokyo University of Agriculture and Technology, Tokyo 184-8588}
\affiliation{CNP, Virginia Polytechnic Institute and State University, Blacksburg, Virginia 24061}
\affiliation{Wayne State University, Detroit, Michigan 48202}
\affiliation{Yamagata University, Yamagata 990-8560}
\affiliation{Yonsei University, Seoul 120-749}
  \author{D.~Liventsev}\affiliation{High Energy Accelerator Research Organization (KEK), Tsukuba 305-0801} 
  \author{I.~Adachi}\affiliation{High Energy Accelerator Research Organization (KEK), Tsukuba 305-0801} 
  \author{H.~Aihara}\affiliation{Department of Physics, University of Tokyo, Tokyo 113-0033} 
  \author{K.~Arinstein}\affiliation{Budker Institute of Nuclear Physics SB RAS and Novosibirsk State University, Novosibirsk 630090} 
  \author{D.~M.~Asner}\affiliation{Pacific Northwest National Laboratory, Richland, Washington 99352} 
  \author{V.~Aulchenko}\affiliation{Budker Institute of Nuclear Physics SB RAS and Novosibirsk State University, Novosibirsk 630090} 
  \author{T.~Aushev}\affiliation{Institute for Theoretical and Experimental Physics, Moscow 117218} 
  \author{A.~M.~Bakich}\affiliation{School of Physics, University of Sydney, NSW 2006} 
  \author{A.~Bay}\affiliation{\'Ecole Polytechnique F\'ed\'erale de Lausanne (EPFL), Lausanne 1015} 
  \author{K.~Belous}\affiliation{Institute for High Energy Physics, Protvino 142281} 
% ...
\collaboration{The Belle Collaboration}

\date{\today}

\begin{abstract}
We report on a search for heavy neutrinos in $B$-meson decays. The
results are obtained using a data sample that contains $772 \times
10^6\, B\bar{B}$ pairs collected at the $\Upsilon(4S)$ resonance with
the Belle detector at the KEKB asymmetric-energy $e^+ e^-$ collider. No
signal is observed and upper limits are set on mixing of heavy neutrinos
with left-handed neutrinos of the Standard Model in the mass range
$0.5~\gevcc-5.0~\gevcc$.

\end{abstract}


\maketitle

{\renewcommand{\thefootnote}{\fnsymbol{footnote}}}
\setcounter{footnote}{0}


The masses of particles in the Standard Model (SM) are generated by the
coupling of the Higgs field to the left- and right-handed components of
a given particle. There being no right-handed neutrino components in the
SM, neutrinos in the SM are strictly massless. However, experimental
data on neutrino oscillations show that neutrinos are not massless,
though their masses are very small~\cite{pdg}. Therefore, a mechanism
beyond the SM is needed to establish neutrino masses. One possibility is
the addition of right-handed neutrinos, which may also have a Majorana
mass, naturally explaining the smallness of the observed neutrino masses
via the so-called ``see-saw'' mechanism~\cite{seesaw}. For example, the
neutrino minimal Standard Model ($\nu$MSM)~\cite{numsm} introduces three
right-handed singlet heavy neutrinos, so that every left-handed particle
has a right-handed counterpart. This model explains neutrino
oscillations, the existence of dark matter and baryogenesis with the
same set of parameters. Heavy neutrinos also appear in other extensions
to the SM, such as SUSY~\cite{susy}, grand unification
theories~\cite{gut} or models with exotic Higgs
representations~\cite{exhiggs}.

\begin{figure*}[tbp]
\begin{minipage}[b]{0.48\textwidth}
\begin{tabular}{c}
\includegraphics[width=0.9\textwidth,clip=true]{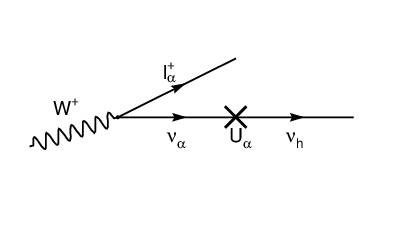}\\
\includegraphics[width=0.9\textwidth,clip=true]{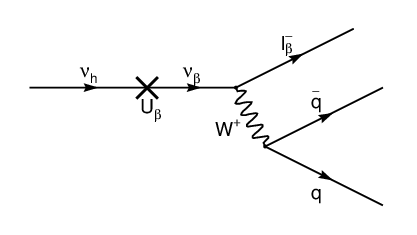}\\
\end{tabular}
\caption{Heavy neutrino production \textit{(top)} and decay \textit{(bottom)} diagrams.}
\label{p:feyn}
\end{minipage}
\hspace{0.02\textwidth}
\begin{minipage}[b]{0.48\textwidth}
\includegraphics[width=\textwidth,clip=true]{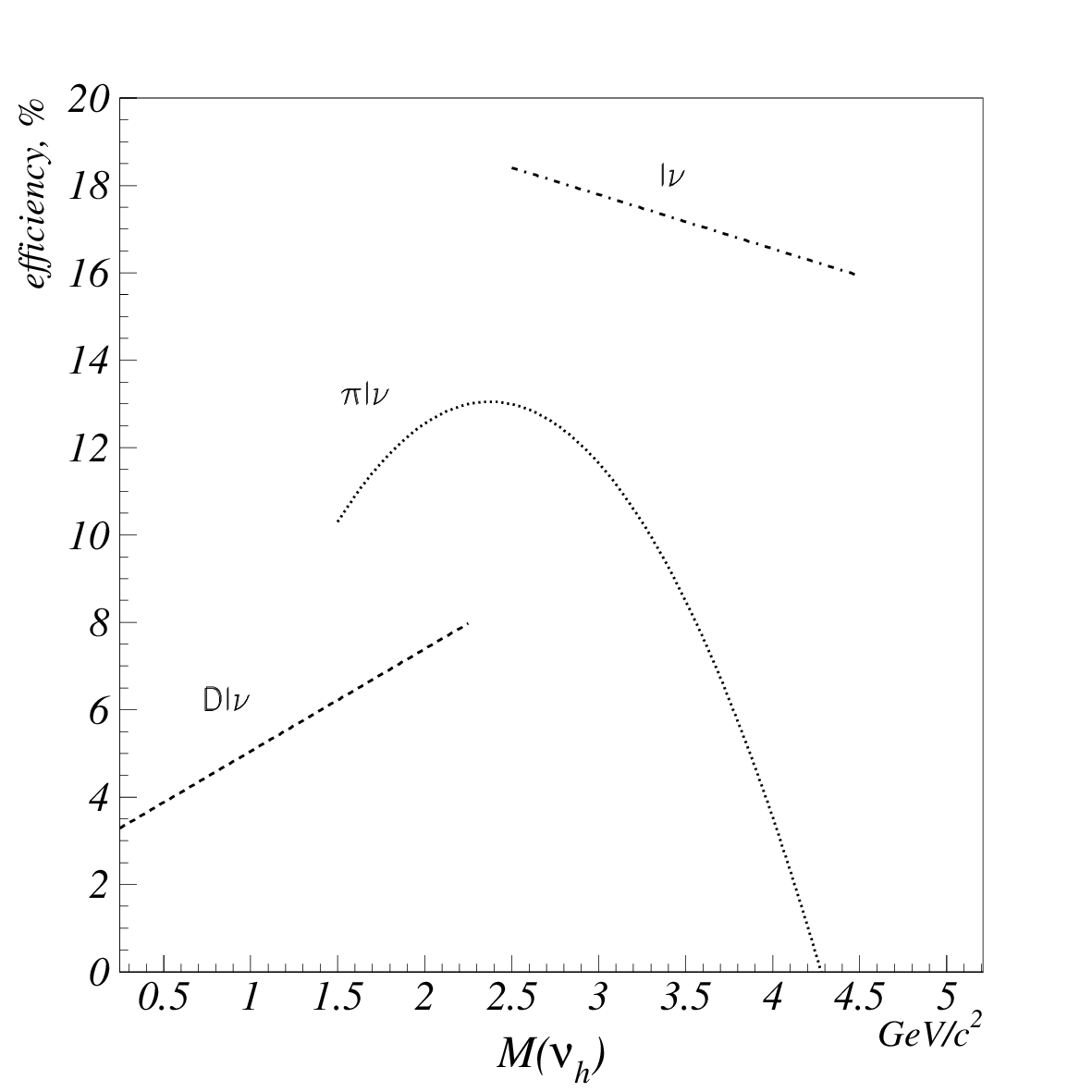}
\caption{Efficiency distributions for different production modes.}
\label{p:eff}
\end{minipage}
\end{figure*}

In general, neutrino flavor eigenstates need not coincide with the mass
eigenstates but may be related through a unitary transformation, similar
to the one that applies to the quark sector,
\begin{equation}
\nu_\alpha = \sum_i{U_{\alpha i}\nu_i}, \quad \alpha = e, \mu, \tau,...,\quad i = 1, 2, 3, 4,...
\end{equation}
where $\alpha$ denotes the flavor eigenstates and $i$ denotes the mass
eigenstates. Production and decay diagrams for heavy neutrinos are shown
in Fig.~\ref{p:feyn}. The coupling of the heavy neutrino $\nu_4$ to the
charged current of flavor $\alpha$ is characterized by a quantity
$U_{\alpha 4}$. Below, we denote a heavy neutrino in the mass range
accessible at Belle and its corresponding coupling constant by $\nu_h$
and $U_{\alpha}$, respectively. Existing experimental results are
reviewed and discussed in Ref.~\cite{exps}.


In this paper, we describe a direct search for heavy neutrino decays
$\nu_h \to \ell^\pm \pi^\mp$, $\ell=e,\mu$ with the Belle detector. The
measurement is based on a data sample that contains 772 million
$B\bar{B}$ pairs, which corresponds to $711\,\mathrm{fb}^{-1}$,
collected at the $\Upsilon(4S)$ resonance with the Belle detector
operating at the KEKB asymmetric-energy $e^+ e^-$ collider~\cite{KEKB}.
The Belle detector is a large-solid-angle magnetic spectrometer that
consists of a silicon vertex detector (SVD), a 50-layer central drift
chamber (CDC), an array of aerogel threshold Cherenkov counters (ACC), a
barrel-like arrangement of time-of-flight scintillation counters (TOF),
and an electromagnetic calorimeter comprised of CsI(Tl) crystals (ECL)
located inside a superconducting solenoid coil that provides a 1.5~T
magnetic field. An iron flux return located outside the coil is
instrumented to detect $K_L^0$ mesons and to identify muons (KLM). The
detector is described in detail elsewhere~\cite{Belle}. Tracking at
Belle is done using the SVD and CDC.

Backgrounds are studied using Monte Carlo (MC) samples of known $B\bar{B}$
decays from $b\to c$ processes (generic MC) that have three times the
statistics of the Belle dataset. Signal MC samples of 500,000 events
each for different heavy neutrino masses and production mechanisms
are used to evaluate the response of the detector, determine its
acceptance and efficiency, and optimize selection criteria. Events are
generated using the EvtGen program~\cite{evtgen}. Heavy neutrinos are
produced and decayed using a phase space model.


At Belle, the most favorable mass range to look for a heavy neutrino is
$M(K)<M(\nu_h)<M(B)$~\cite{gorbunov}. This analysis uses the
leptonic and semileptonic $B$ meson decays $B \to X \ell \nu_h$, where
$\ell=e,\mu$ and $X$ may be a charm meson $D^{(*)}$, a light meson
($\pi$, $\rho$, $\eta$, etc.) or `nothing' (purely leptonic decay), with
relative rates as given in Ref.~\cite{gorbunov}. 

A distinctive feature of the heavy neutrino is its long expected flight
length: for $M(\nu_h)=1\,\gevcc$ and $|U_e|^2=|U_\mu|^2=10^{-4}$ the
flight length is $c\tau \simeq 20$\,m. Therefore, the expected overall
reconstruction efficiency is small. To improve sensitivity, a partial
reconstruction technique is used. A candidate is formed from two
leptons and a pion ($\ell_2 \ell_1 \pi$), where $\ell_1$ and $\pi$ have
opposite charge and form the heavy neutrino candidate with a vertex
displaced from the interaction point (IP). The lepton $\ell_1$ is
referred to as the `signal lepton,' while the lepton $\ell_2$, which
comes from the $B$ decay, is referred to as the `production lepton.' In
this analysis, the heavy neutrino is assumed to be a Majorana fermion and
may decay to a lepton of any charge regardless of the original $B$-meson
flavor. If the heavy neutrino were a Dirac fermion, the production and
decay leptons would necessarily have opposite charge.

If the heavy neutrino is light enough to be produced via $B\to
D^{(*)}\ell\nu_h$, these production modes are expected to dominate over
decays to light mesons due to the small value of the ratio of the
relevant CKM matrix elements $|V_{ub}|/|V_{cb}|$. The background is
more severe for smaller heavy neutrino masses, $M(\nu_h)<2\,\gevcc$, so
an analysis using only $B\to D^{(*)}\ell\nu_h$ modes is used in this
``small mass'' regime, while the full inclusive analysis is used in the
``large mass'' regime.


To suppress the QED background, the charged multiplicity in the event is
required to be larger than four. Charged tracks positively identified as
electrons or muons (as defined in the next paragraph) with
laboratory-frame momentum greater than $0.5\,\mathrm{GeV}/c$ are used as
leptons. All other tracks in the event are treated as pion
candidates. Additional selection criteria for the lepton and pion tracks
are described below.


A significant background remains for heavy neutrino candidates from
particles with similar event topology, notably $K^0_S \to \pi^+ \pi^-$,
$\Lambda \to p \pi^-$, $\gamma \to e^+ e^-$. These backgrounds are
suppressed by strict lepton identification requirements. Electrons are
identified using the energy and shower profile in the ECL, the light
yield in the ACC and the specific ionization energy loss in the CDC
($dE/dx$). This information is used to form an electron
($\mathcal{L}_e$) and non-electron ($\mathcal{L}_{\bar{e}}$)
likelihood; these are combined into a likelihood
ratio $\mathcal{R}_{e} =
\mathcal{L}_{e}/(\mathcal{L}_e+\mathcal{L}_{\bar{e}})$~\cite{eid}. Applying
a requirement on $\mathcal{R}_e$, electrons are selected with an efficiency
and a misidentification rate of approximately 90\% and 0.1\%,
respectively, in the kinematic region of interest. Muons are
distinguished from other charged tracks by their range and hit profiles
in the KLM. This information is utilized in a likelihood ratio
approach~\cite{muid} similar to the one used for the electron
identification. Muons are selected with an efficiency and a
misidentification rate of approximately 90\% and 1\%, respectively, in
the kinematic region of interest. These requirements are reversed in
order to produce a lepton veto for identifying pion candidates.


We select well-vertexed heavy neutrino candidates using $dr$, the
distance of closest approach to the IP in the plane perpendicular to the
beam axis for each track; $d\phi$, the angle between the momentum vector
and decay vertex vector of the heavy neutrino candidate; and
$dz_\textrm{vtx}$, the distance between the daughter tracks at their
closest approach in the direction parallel to the beam. Requirements
vary depending on the presence of SVD hits on the tracks and on the
heavy neutrino candidate flight length.
The signal lepton and pion are fit to a common
vertex. Only candidates with $\chi_1^2/ndf<16$, where $\chi_1^2$ is the
goodness of fit and $ndf$ is the number of degrees of freedom, are
accepted. A second vertex fit of the heavy neutrino candidate and the
production lepton is performed with the vertex constrained to the IP;
candidates with $\chi_2^2/ndf<4$ are retained.


For combinatorial background, the daughter tracks of the heavy neutrino
candidate often originate from the vicinity of the IP rather than the
candidate's decay vertex. In order to suppress this background, the
difference between the radial coordinates of the closest associated hit
in the SVD or CDC of either of the two daughter tracks to the IP
($r_{\ell}$ or $r_{\pi}$) and the candidate's decay vertex
($r_\textrm{vtx}$) is calculated:
$dr_\textrm{fh}=\min(r_\ell,r_\pi)-r_\textrm{vtx}$. This requirement is
most effective for large $r_\textrm{vtx}$. The analysis requires
$dr_\textrm{fh}>-2$\,cm for $r_\textrm{vtx}>6\,\textrm{cm}$.


For the ``small mass'' ($M(\nu_h)<2\,\gevcc$) analysis, $B \to
D^{(*)}\ell\nu_h$ events are selected using the recoil mass against the
$\ell\ell\pi$ system. This requirement is related to the kinematics of
the decay under study. For $B \to X \ell \nu_h \to X \ell \ell \pi$
decays, the mass of $X$ can be obtained from $M_X^2 =
(E_{CM}-E_{\ell\ell\pi})^2 - P_{\ell\ell\pi}^2 - P_B^2 - 2 \vec
P_{\ell\ell\pi} \cdot \vec P_B$, where $E_\textrm{CM}$ and $P_B$ are the
$B$ meson center-of-mass (CM) energy and momentum and $E_{\ell\ell\pi}$
and $P_{\ell\ell\pi}$ are the CM energy and momentum of the
$\ell\ell\pi$ system. The last term in this equation cannot be
calculated as the $B$ direction remains unknown, so we redefine the
recoil mass as $M_X^2 \equiv (E_\textrm{CM}-E_{\ell\ell\pi})^2 -
P_{\ell\ell\pi}^2 - P_B^2$. For events with $X=D^{(*)}$, the $M_X$
distribution has overlapping peaks around the masses of the $D$ and
$D^*$, while for background events the recoil mass has a broader
distribution. Events with
$1.4\,\gevcc<M_X<2.4\,\gevcc$ are selected as candidates.

To reject protons from the decays of long-lived baryons, we impose a
loose proton veto for the pion candidate. For each track, the likelihood
values $\mathcal{L}_p$ and $\mathcal{L}_K$ of the proton and kaon
hypotheses, respectively, are determined from the information provided
by the hadron identification system (CDC, ACC, and TOF). A track is
identified as a proton if
$\mathcal{L}_p/(\mathcal{L}_p+\mathcal{L}_K)>0.99$. Background events,
rejected by the veto, are concentrated at heavy neutrino masses below
$2\,\gevcc$ and thus this veto is applied in the ``small mass'' analysis
only.

\begin{figure*}[tbp]
\begin{minipage}[b]{0.98\textwidth}
\begin{tabular}{cc}
\includegraphics[width=0.5\textwidth,clip=true]{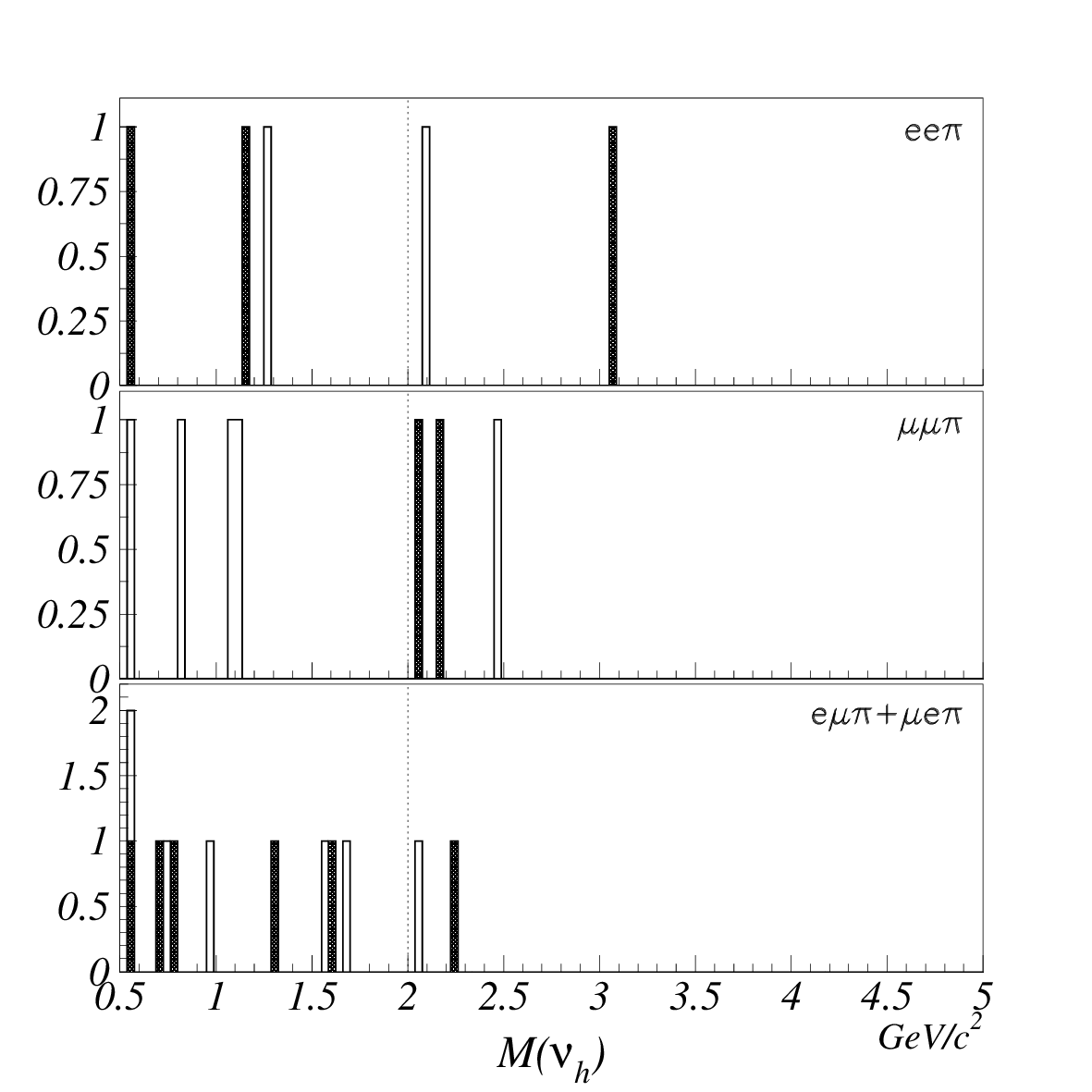}&
\includegraphics[width=0.5\textwidth,clip=true]{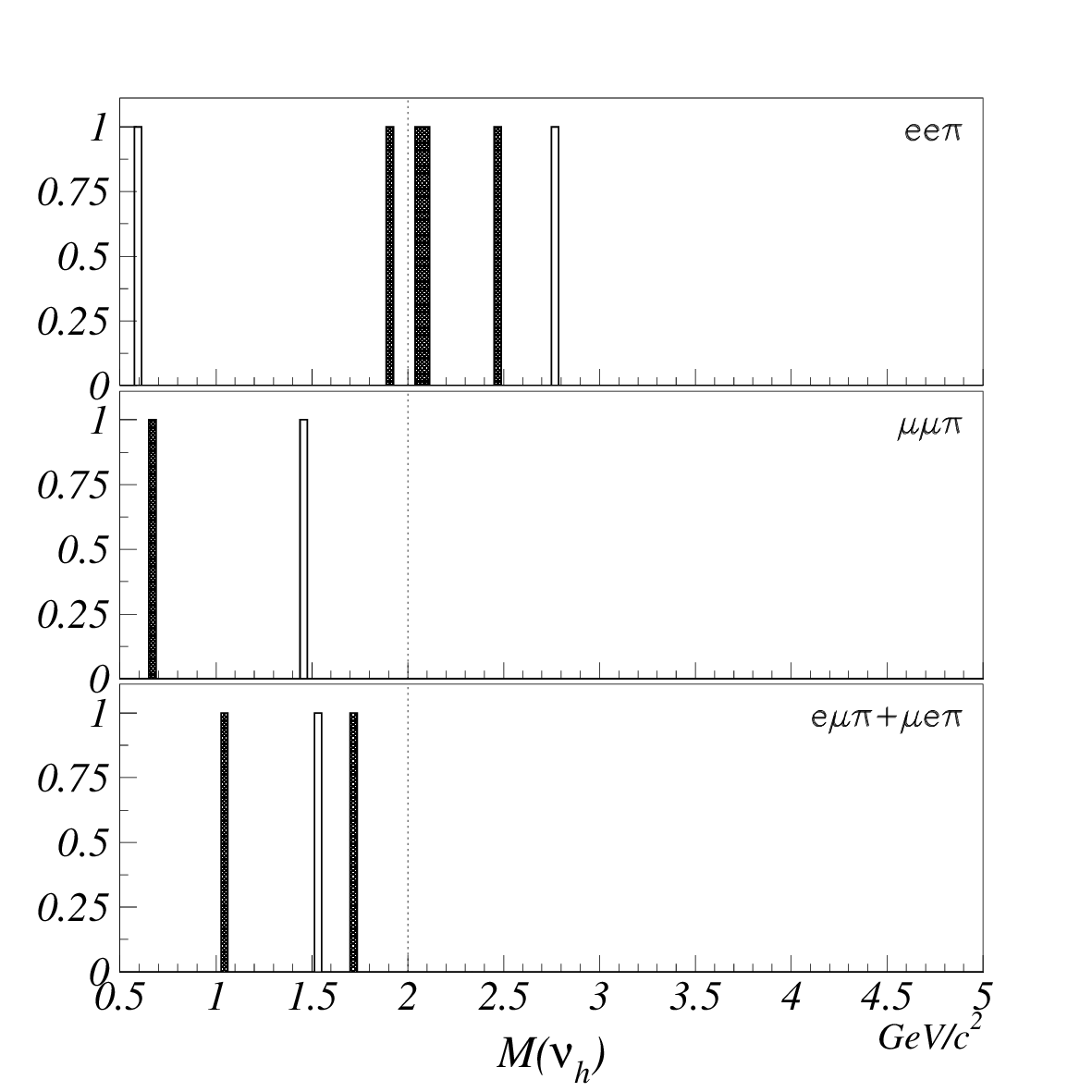}\\
\textit{a)} & \textit{b)} \\
\end{tabular}
\caption{Distributions of $M(\nu_h)$ for $ee\pi$, $\mu\mu\pi$ and
  $e\mu\pi+\mu e\pi$ reconstruction modes in generic MC (unscaled)
  \textit{(a)}, and data \textit{(b)}. The dotted line shows the
  boundary between the ``small mass'' and ``large mass'' methods. The
  filled (black) histograms are for candidates with opposite-charge
  leptons, while the open (white) histograms are for candidates with
  same-charge leptons.}
\label{p:res_ev}
\end{minipage}
\end{figure*}

Using the requirements described above, the number of background events
is reduced by a factor of $\sim 10^6$ to a handful of events, as shown
in Fig.~\ref{p:res_ev}. Their summary is shown in Table~\ref{t:cutsum}.
The five event types in the Table are: I: both neutrino daughter tracks
have recorded hits in SVD, II: one of the neutrino daughter tracks has
recorded hits in SVD, III: none of the neutrino daughter tracks have
recorded hits in SVD, and $r_\textrm{vtx}<12\,\textrm{cm}$, IV: no SVD
hits and $12\,\textrm{cm}<r_\textrm{vtx}<30\,\textrm{cm}$, V: no SVD
hits and decay radius exceeds $r_\textrm{vtx}>30\,\textrm{cm}$. The
reconstruction efficiency for signal events does not depend
significantly on the reconstruction mode ($ee\pi$, $\mu\mu\pi$ or
$e\mu\pi$), but does depend strongly on the heavy neutrino mass. For a
given mass, the efficiency also depends on the $B$-meson decay mode in
which the heavy neutrino is produced.  Efficiency distributions,
including reconstruction efficiency, for different production modes are
shown in Fig.~\ref{p:eff}. Efficiency of the requirements alone does not
depend much on mass or production mode. Table~\ref{t:cutsum} shows
requirements efficiency for $D\ell\nu_h$ mode and
$M(\nu_h)=2\gevcc$. The efficiency drops with the radius
$r_\textrm{vtx}$ of the decay vertex from the beam axis. The effective
range of neutrino reconstruction extends to $r_\textrm{vtx} \simeq
60\,\textrm{cm}$.

\begin{table}[htp]
\caption{Summary of requirements, their background suppression
  efficiency, efficiency for signal events and systematic
  uncertainties.}
\label{t:cutsum}
\begin{center}
\begin{tabular}{c|c|c|c|c}
Requirement & Applied      & Supp.    & Signal   & Syst. \\
            & to           & eff., \% & eff., \% & error, \% \\
\hline \hline
$\chi_1^2/ndf<16$             & All & 35 & 99 & 2.9 \\
$\chi_2^2/ndf<4$             & All & 27 & 85 & 10.1 \\
${\cal R}_e(\ell_1)>0.9$         & All & 40 & 45 & 2.2 \\
${\cal R}_\mu(\ell_1)>0.99$      & All & 17 & 35 & 4.9 \\
${\cal R}_e(\ell_2)>0.9$         & All & 38 & 53 & 3.0 \\
${\cal R}_\mu(\ell_2)>0.9$       & All & 25 & 38 & 3.1 \\
Lepton veto            & All & 86 & 99 & 1.8 \\
\hline
$d\phi<0.03\,\textrm{cm}$       & Type I & 39 & 95 & \multirow{5}{*}{$\left.\rule{0cm}{1.2cm}\right\}$5.8} \\
$d\phi<0.03\,\textrm{cm}$       & Type II &  5 & 80 & \\
$d\phi<0.04\,\textrm{cm}$       & Type III & 11 & 85 & \\
$d\phi<0.09\,\textrm{cm}$       & Type IV & 66 & 96 & \\
$d\phi<0.15\,\textrm{cm}$       & Type V & 51 & 94 & \\
\hline
$dr>0.09\,\textrm{cm}$          & Type I &  5 & 97 & \multirow{5}{*}{$\left.\rule{0cm}{1.2cm}\right\}$3.7} \\
$dr>0.1\,\textrm{cm}$          & Type II &  7 & 98 & \\
$dr>3\,\textrm{cm}$          & Type III &  1 & 79 & \\
$dr>3\,\textrm{cm}$          & Type IV & 10 & 94 & \\
$dr>5\,\textrm{cm}$          & Type V & 42 & 95 & \\
\hline
$dz_\textrm{vtx}<0.4\,\textrm{cm}$       & Type I & 37 & 94 & \multirow{5}{*}{$\left.\rule{0cm}{1.2cm}\right\}$10.0} \\
$dz_\textrm{vtx}<0.4\,\textrm{cm}$       & Type II & 17 & 74 & \\
$dz_\textrm{vtx}<0.5\,\textrm{cm}$       & Type III & 21 & 75 & \\
$dz_\textrm{vtx}<0.9\,\textrm{cm}$       & Type IV & 36 & 80 & \\
$dz_\textrm{vtx}<2\,\textrm{cm}$       & Type V & 68 & 83 & \\
\hline
$dr_\textrm{fh}>-2\,\textrm{cm}$  & $r_\textrm{vtx}>6\,\textrm{cm}$ & 32 & 84 & 2.9 \\
Recoil mass           & Small mass & 24 & 99 & 4.1 \\
Proton veto           & Small mass & 94 & 97 & 1.6 \\
\end{tabular}
\end{center}
\end{table}


If the heavy neutrino lifetime is long enough, then the number of
neutrinos detected in the Belle detector is (in units where $\hbar=c=1$)
\begin{widetext}
\begin{align}
\label{calc}
n(\nu_h) & = 2N_{BB}\ \mathcal{B}(B \to X \ell \nu_h)\ \mathcal{B}(\nu_h \to \ell \pi)\ \int \varepsilon(R) \frac{m\Gamma}{p} \exp\Big(-\frac{mR\Gamma}{p}\Big) dR\nonumber\\
         & \simeq |U_\alpha|^2|U_\beta|^2\ 2N_{BB}\ f_1(m)\ f_2(m)\ \frac{m}{p} \int \varepsilon(R) dR,
\end{align}
\end{widetext}
where $N_{BB}$ is the number of $B\bar{B}$ pairs, $\mathcal{B}(B \to X
\ell \nu_h)$ is the total branching fraction for $\nu_h$ production,
$\mathcal{B}(\nu_h \to \ell \pi)$ is the branching fraction of the
reconstructed decay, $\varepsilon(R)$ is the reconstruction efficiency
of the $\nu_h$ decaying at a distance $R$ from the IP and $m$, $p$ and
$\Gamma$ are the mass, momentum and full width of the heavy neutrino,
respectively. Additionally, to factor out the $|U|^2$ dependence, we
define $|U_\alpha|^2f_1(m)\equiv\mathcal{B}(B \to X \ell \nu_h)$ and
$|U_\beta|^2f_2(m)\equiv\Gamma(\nu_h \to \ell \pi) = \mathcal{B}(\nu_h
\to \ell \pi) \Gamma$, where $\alpha$ and $\beta$ denote the flavor of
the charged lepton produced in the $B$ and $\nu_h$ decay,
respectively. The exponent in the integrand of Eqn.~\eqref{calc} is
approximated by unity. An error introduced by this approximation is
small and is negligible when the flight length is long enough
(for $|U|^2\lesssim 10^{-3}$). Integration is performed over the full
volume used to reconstruct the heavy neutrino vertex, which depends on
the reconstruction requirements. The expressions for $\mathcal{B}(B \to
X \ell \nu_h)$ and $\Gamma(\nu_h \to \ell \pi)$ are taken from
Ref.~\cite{gorbunov} and require only very general assumptions (i.e.,
they are not specific to $\nu$MSM).

The calculated total branching fractions for heavy neutrino production
$\mathcal{B}(B \to X \ell \nu_h)$ for the ``small mass'' and ``large
mass'' analyses correspond to
\begin{equation}
\mathcal{B}(B \to X \ell \nu_h)_\textrm{small mass}=\mathcal{B}(B \to D \ell \nu_h)+\mathcal{B}(B \to D^* \ell \nu_h)
\end{equation}
and
\begin{equation}
\mathcal{B}(B \to X \ell \nu_h)_\textrm{large mass}=\sum_i \mathcal{B}(B \to X_i \ell \nu_h),
\end{equation}
respectively, where the summation is done over $D$, $D^*$, $\pi$,
$\rho$, $\eta$, $\eta'$, $\omega$, $\phi$ and `nothing.' These are not
exact expressions but rather estimates of lower bounds on $\mathcal{B}(B
\to X \ell \nu_h)$, which lead to conservative upper limits on $|U|^2$.


The systematic uncertainty of each of the event selection criteria is
estimated from the difference in the efficiencies obtained in data and
MC. A summary of all systematic uncertainties is presented in
Table~\ref{t:cutsum}. Since all particles used in the systematic
uncertainty study decay relatively close to the IP compared to the
expectation for a heavy neutrino, we require where possible that the
decay vertices be farther than 4\,cm from the IP in the transverse plane
to put more weight on large decay lengths. To estimate the systematic
uncertainty due to tracking, we compare the number of fully and
partially reconstructed $D^*$ decays in the decay chain $D^* \to D
\pi^+$, $D \to K_S^0 \pi \pi$, $K_S^0 \to \pi\pi$, where in the latter
case one of the pions from the $K_S^0$ is explicitly left
unreconstructed. To estimate the systematic uncertainty of the recoil
mass requirement, we reconstruct $B \to D D_s^{(*)}$, $D \to K_S^0 \pi$
events and study the mass recoiling against the $D$-meson. The $D$ decay
topology is similar to $\ell \nu_h$ here, and we treat the difference in
recoil mass efficiency between data and MC as the systematic uncertainty
of the recoil mass requirement. To estimate the systematic uncertainty
of the electron identification, we reconstruct $\pi^0\to\gamma\gamma$
events, where one of the photons converts into $e^+e^-$ in the detector
and one of these conversion particles is identified as an electron. The
difference of the identification efficiency of the other daughter
between data and MC is treated as a systematic uncertainty. For the muon
identification, we perform a similar study with a $J/\psi\to\mu^+\mu^-$
sample. To estimate the systematic uncertainty of other reconstruction
requirements, we apply these requirements to $K^0_S$ decays, which have
a topology similar to heavy neutrino decays. Correlations between
different systematic uncertainties are found to be small and are
neglected. All systematic uncertainties are summed in quadrature,
leading to total systematic uncertainties of 25.0\% and 25.4\% for the
``small mass'' and ``large mass'' regimes, respectively. The largest
contributions to the systematic uncertainties are $\chi_2^2$ (10.1\%),
$dz_\textrm{vtx}$ (10.0\%) and tracking of the heavy neutrino candidate
daughter particles (8.7\% per track, added linearly).

\begin{figure*}[tbp]
\begin{minipage}[b]{0.98\textwidth}
\begin{tabular}{cc}
\includegraphics[width=0.5\textwidth]{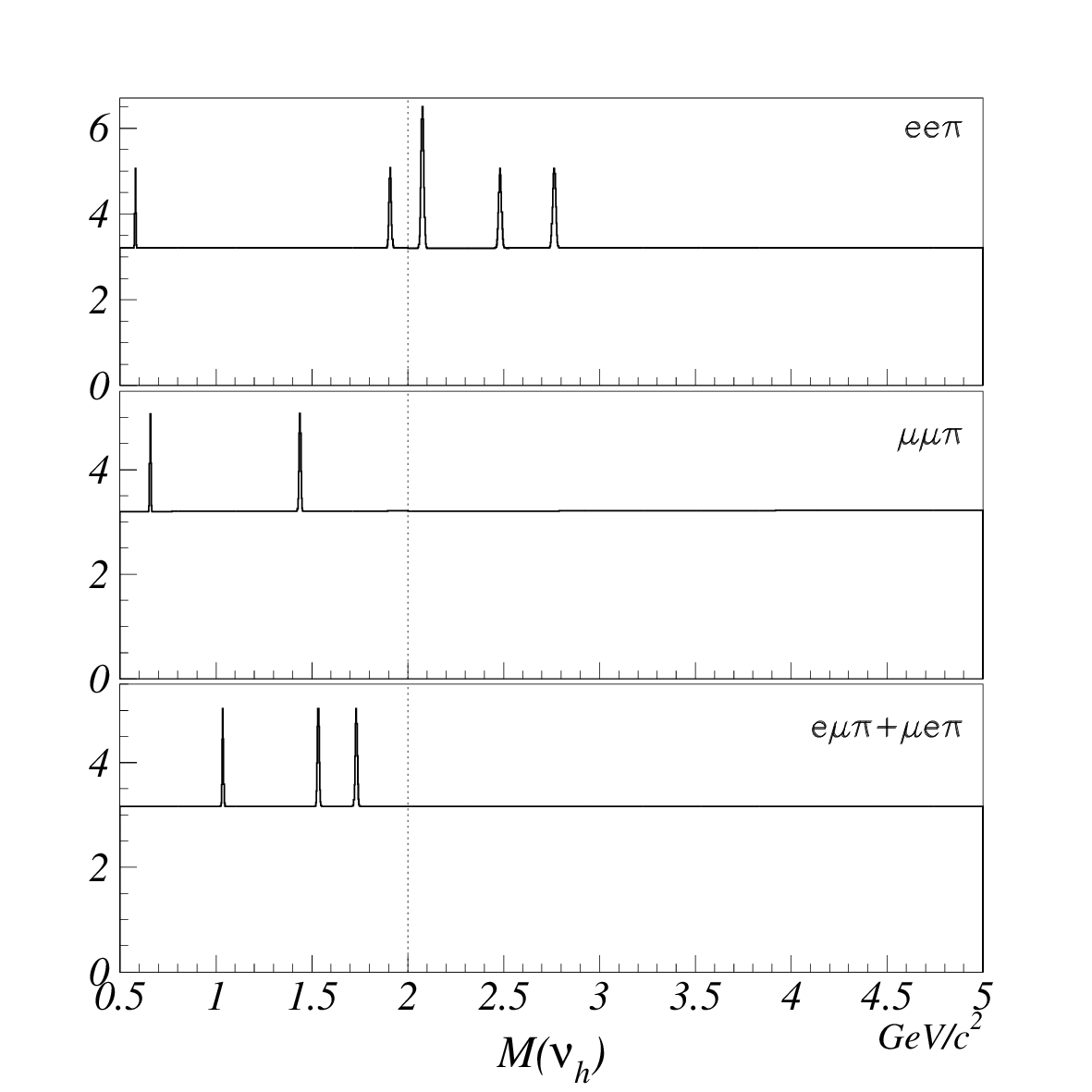}&
\includegraphics[width=0.5\textwidth,clip=true]{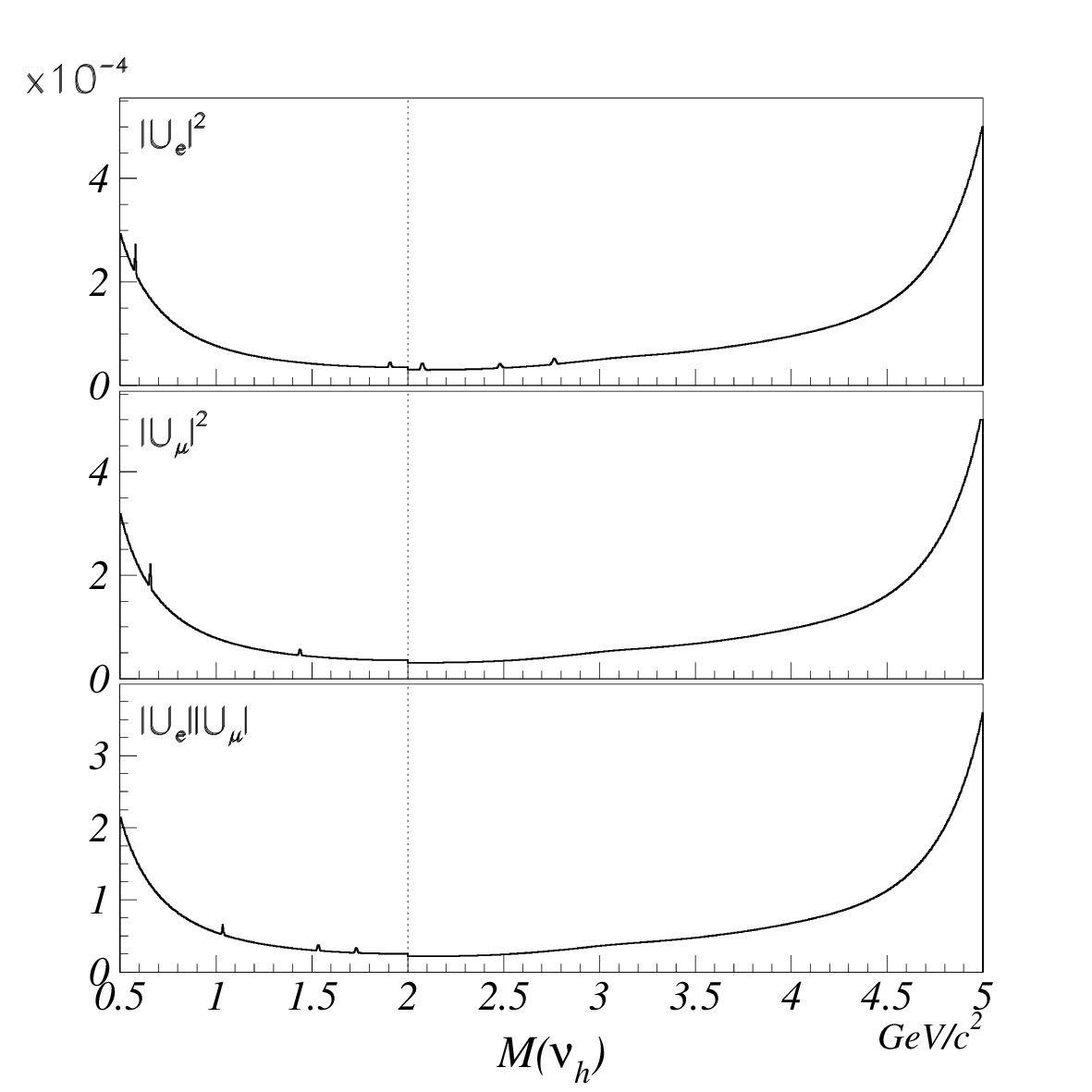}\\
\textit{a)} & \textit{b)} \\
\end{tabular}
\caption{Upper limits at 90\% CL on the number of signal events
  \textit{(a)} and $|U_e|^2$, $|U_\mu|^2$ and $|U_e||U_\mu|$
  \textit{(b)}. The dotted line shows the boundary between the ``small
  mass'' and ``large mass'' methods.}
\label{p:res_uplim}
\end{minipage}
\end{figure*}

After all the event selection criteria were fixed from the MC study, the
data were analyzed and the coupling constants $|U_e|^2$, $|U_\mu|^2$ and
$|U_e||U_\mu|$ were obtained separately using the decay modes $ee\pi$,
$\mu\mu\pi$ and $e\mu\pi+\mu e\pi$, respectively. Distributions of the
heavy neutrino mass in generic MC and data are shown in
Fig.~\ref{p:res_ev}. In agreement with MC expectations, only a few
isolated events are observed and we set upper limits on $|U|^2$
according to Ref.~\cite{feldman}, taking into account the systematic
uncertainty calculated above. For non-empty bins and empty bins far from
non-empty bins, we set Poisson upper limits, assuming small background,
as suggested from the MC study. In the vicinity of non-empty bins, we
use Gaussian fits to interpolate between empty and non-empty
regions. The widths of the Gaussians are fixed from MC. We use bins of
$3\,\mev/c^2$ width, since the mass resolution evolves from $\sim
3\,\mev/c^2$ at $M(\nu_h)=1\,\gevcc$ to $\sim 12\,\mev/c^2$ at
$M(\nu_h)=4\,\gevcc$. The resulting upper limits at 90\%~CL on the
number of events and coupling constants are shown in
Fig.~\ref{p:res_uplim}.


In conclusion, upper limits on the mixing of heavy right-handed
neutrinos with the conventional SM left-handed neutrinos in the mass
range $0.5-5.0\,\gevcc$ have been obtained. The maximum sensitivities
are achieved around $2\,\gevcc$ and are $3.0\times10^{-5}$,
$3.0\times10^{-5}$ and $2.1\times10^{-5}$ for $|U_e|^2$, $|U_\mu|^2$ and
$|U_e||U_\mu|$, respectively. The corresponding upper limit for the
product branching fraction is $\mathcal{B}(B\to X\ell\nu_h
)\times\mathcal{B}(\nu_h\to\ell\pi^+)<7.2\times10^{-7}$ for $\ell=e$ or
$\mu$. A comparison with existing results for $|U_e|^2$ and $|U_\mu|^2$
is shown in Fig.~\ref{p:comp}.

We thank the KEKB group for the excellent operation of the accelerator;
the KEK cryogenics group for the efficient operation of the solenoid;
and the KEK computer group, the National Institute of Informatics, and
the PNNL/EMSL computing group for valuable computing and SINET4 network
support. We acknowledge support from the Ministry of Education,
Culture, Sports, Science, and Technology (MEXT) of Japan, the Japan
Society for the Promotion of Science (JSPS), and the Tau-Lepton Physics
Research Center of Nagoya University; the Australian Research Council
and the Australian Department of Industry, Innovation, Science and
Research; the National Natural Science Foundation of China under
contract No.~10575109, 10775142, 10875115 and 10825524; the Ministry of
Education, Youth and Sports of the Czech Republic under contract
No.~LA10033 and MSM0021620859; the Department of Science and Technology
of India; the Istituto Nazionale di Fisica Nucleare of Italy; the BK21
and WCU program of the Ministry Education Science and Technology,
National Research Foundation of Korea Grant No.\ 2010-0021174,
2011-0029457, 2012-0008143, 2012R1A1A2008330, BRL program under NRF
Grant No. KRF-2011-0020333, and GSDC of the Korea Institute of Science
and Technology Information; the Polish Ministry of Science and Higher
Education and the National Science Center; the Ministry of Education and
Science of the Russian Federation and the Russian Federal Agency for
Atomic Energy; the Slovenian Research Agency; the Swiss National Science
Foundation; the National Science Council and the Ministry of Education
of Taiwan; and the U.S.\ Department of Energy and the National Science
Foundation. This work is supported by a Grant-in-Aid from MEXT for
Science Research in a Priority Area (``New Development of Flavor
Physics''), and from JSPS for Creative Scientific Research (``Evolution
of Tau-lepton Physics'').


\begin{figure*}[tbp]
\begin{minipage}[b]{0.98\textwidth}
\begin{tabular}{cc}
\includegraphics[width=0.5\textwidth,clip=true]{paper5a.eps}&
\includegraphics[width=0.5\textwidth,clip=true]{paper5b.eps}\\
\textit{a)} & \textit{b)} \\
\end{tabular}
\caption{Comparison of the obtained upper limits for $|U_e|^2$
  \textit{(a)} and $|U_\mu|^2$ \textit{(b)} with existing experimental
  results from CHARM~\cite{charm}, CHARMII~\cite{charmii},
  DELPHI~\cite{delphi}, NuTeV~\cite{nutev}, BEBC~\cite{bebc} and
  NA3~\cite{na3}.}
\label{p:comp}
\end{minipage}
\end{figure*}

\end{document}




\title{Erratum: Search for heavy neutrinos at Belle [Phys. Rev. D 87, 071102(R) (2013)]}

\affiliation{CNP, Virginia Polytechnic Institute and State University, Blacksburg, Virginia 24061}
\affiliation{High Energy Accelerator Research Organization (KEK), Tsukuba 305-0801}
\author{D.~Liventsev}\affiliation{CNP, Virginia Polytechnic Institute and State University, Blacksburg, Virginia 24061}\affiliation{High Energy Accelerator Research Organization (KEK), Tsukuba 305-0801} 
\collaboration{The Belle Collaboration}

\date{\today}

\maketitle

{\renewcommand{\thefootnote}{\fnsymbol{footnote}}}
\setcounter{footnote}{0}

The number of neutrinos detected in the Belle detector is (Eqn.~(2) of Ref.~\cite{orig}):
\begin{widetext}
\begin{align}
n(\nu_h) & = 2N_{BB}\ \mathcal{B}(B \to \nu_h)\ \mathcal{B}(\nu_h \to \ell \pi) \int \frac{m\Gamma}{p} \exp\Big(-\frac{m\Gamma R}{p}\Big) \varepsilon(R) dR\\
         & = |U_\alpha|^2|U_\beta|^2\ 2N_{BB}\ f_1(m)\ f_2(m)\ \frac{m}{p} \int \exp\Big(-\frac{m\Gamma R}{p}\Big) \varepsilon(R) dR\label{e:nnufull}
\end{align}
\end{widetext}
where $N_{BB}$ is the number of $B\bar{B}$ pairs, $\mathcal{B}(B \to
\nu_h)$ is the total branching fraction for $\nu_h$ production,
$\mathcal{B}(\nu_h \to \ell \pi)$ is the branching fraction of the
reconstructed decay, $\varepsilon(R)$ is the reconstruction efficiency
of the $\nu_h$ decaying at a distance $R$ from the interaction point,
$m$, $p$ and $\Gamma$ are the mass, momentum and full width of the heavy
neutrino, respectively. Integration is performed over the full volume
used to reconstruct the heavy neutrino vertex, which depends on
reconstruction requirements. Additionally, to factor out the $|U|^2$
dependence, we define $|U_\alpha|^2f_1(m) \equiv \mathcal{B}(B \to
\nu_h)$ and $|U_\beta|^2f_2(m) \equiv \Gamma(\nu_h \to \ell \pi) =
\mathcal{B}(\nu_h \to \ell \pi) \Gamma$, where $\alpha$ and $\beta$
denote the flavor of the charged lepton produced in the $B$ and $\nu_h$
decays, respectively.

In the original paper~\cite{orig}, the exponent in the integrand was
approximated by unity under the assumption that $\Gamma$ is small and
the flight length is sufficiently long:
\begin{widetext}
\begin{equation}
n(\nu_h) \simeq |U_\alpha|^2|U_\beta|^2\ 2N_{BB}\ f_1(m)\ f_2(m)\ \frac{m}{p} \int \varepsilon(R) dR.\label{e:nnureduced}
\end{equation}
\end{widetext}
Here, the momentum of the heavy neutrino was kept constant.

The assumption of a long flight length is valid only for a small
neutrino mass. However, since the partial decay widths approximately
increase with mass as $m^3$ to $m^5$ depending on the final
state~\cite{gorbunov}, at higher masses this approximation introduces a
significant overestimation of the upper limit on $|U|^2$. Moreover, at
large masses there is also an upper bound of the excluded region since,
for some values of the coupling constant, the flight length is so small
that the event is rejected by the selection criteria. Therefore, we
recalculate the upper limits on $|U|^2$ taking into account the exponent
in Eqn.~\ref{e:nnufull} and the actual neutrino momentum. The full
neutrino width $\Gamma$ is calculated as the sum of partial widths from
Ref.~\cite{gorbunov}. Since we do not know the relation among the
different $U_\alpha$ we assume $|U|^2 = |U_\alpha||U_\beta|$ in the
calculation of $\Gamma$. To obtain the upper limit on the coupling
constants, we solve the Eqn.~\ref{e:nnufull} for the variable $|U|^2$.

\begin{figure*}[htbp]
\begin{minipage}[b]{0.48\textwidth}
\begin{tabular}{c}
\includegraphics[width=1.2\textwidth]{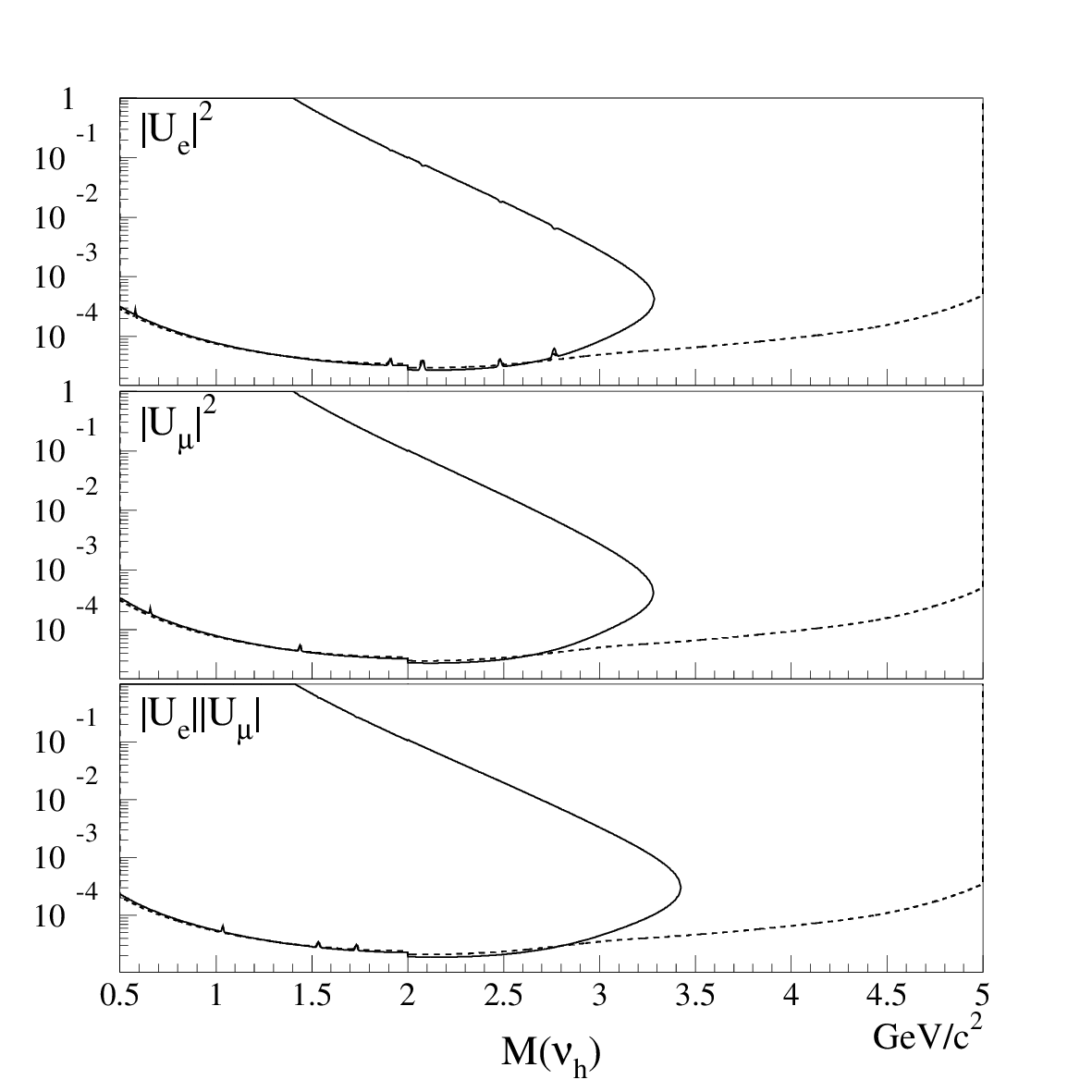} 
\end{tabular}
\caption{Upper limits at 90\% CL on the $|U_e|^2$, $|U_\mu|^2$ and
  $|U_e||U_\mu|$ The solid line shows the correct exclusion region; the
  dashed line shows the one from the original paper.}
\label{p:res_uplim}
\end{minipage}
\end{figure*}

The updated versions of Figs.~4b and 5 in the original paper with the
new calculation are shown here as Figs.~\ref{p:res_uplim} and
\ref{p:comp}, respectively.

We thank Brian Shuve and Michael Peskin for pointing out the issue.


\begin{figure*}[tbp]
\begin{minipage}[b]{0.98\textwidth}
\begin{tabular}{cc}
\includegraphics[width=0.5\textwidth,clip=true]{fig2a.eps}&
\includegraphics[width=0.5\textwidth,clip=true]{fig2b.eps}\\
\textit{a)} & \textit{b)} \\
\end{tabular}
\caption{Comparison of the obtained upper limits for $|U_e|^2$
  \textit{(a)} and $|U_\mu|^2$ \textit{(b)} with existing experimental
  results from CHARM, CHARMII, DELPHI, NuTeV, BEBC and NA3.}
\label{p:comp}
\end{minipage}
\end{figure*}